\documentclass[a4paper]{article}

\usepackage{icrc2013}
\usepackage[english]{babel}
\usepackage{lineno}

\title{An outdoor test facility for the Cherenkov Telescopes Array mirrors}

\shorttitle{Outdoor test facility}

\authors{
Medina,~M.~C. $^{1}$,
Garc\'ia,~B.$^{2}$,
Maya,~J.$^{2}$,
Mancilla,~A.$^{2}$,
Larrarte,~J.~J.$^{1}$,
Rasztocky,~E.$^{1}$,
Benitez,~M.$^{1}$,
Dipold,~J.$^{3}$ and
Platino, M.$^{4}$
for the CTA Consortium$^{5}$.
}

\afiliations{
$^1$ Instituto Argentino de Radioastronomia, CCT La Plata-CONICET, Argentina \\
$^2$ ITeDA, Mendoza, Argentina \\
$^3$ IFSC, USP, Brazil \\
$^4$ ITeDA, Buenos Aires, Argentina \\
$^5$ www.cta-observatory.org
}

\email{clementina@iar.unlp.edu.ar}

\abstract{The Cherenkov Telescopes Array (CTA) is planned to be an Observatory for very high energy gamma ray astronomy and will consist of several tens of telescopes which account for a reflective surface of more than 10000 m$^2$. The mirrors of these telescopes will be formed by a set of facets. Different technological solutions, for a fast and cost efficient production of light-weight mirror facets are under test inside the CTA Consortium. Most of them involve composite structures whose behavior under real
observing conditions is not yet fully tested. An outdoor test facility has been built in one of the candidate sites for CTA, in Argentina (San Antonio de los Cobres [SAC], 3600m a.s.l) in order to monitor the optical and mechanical properties of these facets exposed to the local atmospheric conditions for a given period of time. In this work we present the preliminary results of the first Middle Size Telescope (MST) mirror-monitoring campaign, started in 2013.}


\begin{document}

\maketitle

\section{Introduction}

The next generation of Very High Energy (VHE) $\gamma$-ray telescope array is CTA (Cherenkov Telescope Array), which is currently in the development phase\cite{bib:CTA}. Two sites, one in the Northern and one in the Southern Hemisphere, are planned to provide full-sky coverage. In each of these sites an array of telescopes of multiple sizes will be installed; there will be small (5~m), medium (12~m) and large (23~m) diameter telescopes (called, from now on, Small Size Telescope or SST, Medium Size Telescope or MST and Large Size Telescope or LST, respectively), each optimized for different energy ranges.

The final configurations of these arrays are not yet completely defined but the southern site of CTA will be composed of at least 50 telescopes of 3 different sizes and a total of over 5,000~m$^2$ of mirrors will be necessary. The northern site, which is intended to be smaller, will require of the order of 3,500~m$^2$ of mirrors. Because of its large size, the reflector of a Cherenkov telescope is composed of many individual mirror facets. In particular, for the MST, hexagonal mirrors of 1.2\,m (flat to flat) diameter will be used, with a spherical shape of about 32\,m of radius of curvature. In order to fulfill the specifications on optical properties, mechanical behavior and costs, different technological solutions are under study \cite{bib:CTA}\cite{bib:CTAMir}. Most of them involve a composite structure, supporting a slim reflective surface, which is assembled and glued using the cold slump technique \cite{bib:MagicMir2}\cite{bib:Saclay}.  

In this work we present the first outdoor test facility for the CTA composite mirrors. This facility is placed in one of the Argentinean candidate sites for the southern observatory and it will allow the characterization of the behavior of different mirror designs under real environmental conditions. The paper is organized as follows: in section \ref{sec:mir} the mirrors under test are described. In section~\ref{sec:site} a small description of the site is given. Section~\ref{sec:facility} is dedicated to the description of the facility and tests, while in section~\ref{sec:results} some very preliminary results are presented. Finally, section~\ref{sec:conc} contains the conclusions and the perspectives of this work. 
 
\section{Mirrors for Cherenkov telescopes}\label{sec:mir}

\begin{figure}[h]
  \centering
  \includegraphics[width=0.45\textwidth] {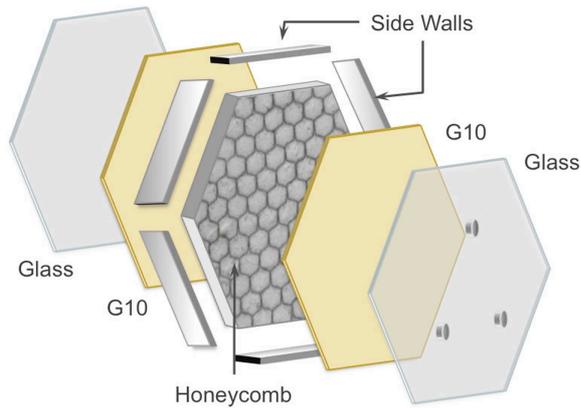}
  \caption{Exploded view of the Irfu-CEA and Kerdry mirrors.}
  \label{fig:mirrors}
 \end{figure}

The Irfu-CEA CTA team, together with the Kerdry company provided two MST mirrors manufactured as described in \cite{bib:Saclay} to be tested with the facility installed at San Antonio de los Cobres (SAC, from now on). The most important characteristics of these particular mirrors are described in Table \ref{table:mirrors}. Schematics of the inner structure of the mirrors is shown in Fig.\ref{fig:mirrors}. They are part of the first series of prototypes intended to be used for testing the behavior of the mirrors against different typical parameters on several astronomical sites, i.~e. extreme temperature gradients, humidity levels, atmospheric dust, etc.

\begin{table}[h]
\begin{center}
\begin{tabular}{|c|c|c|c|}
\hline {\tiny Mirror} &  {\tiny Focal length (m)} & {\tiny Reflectivity ($\%$)} & {\tiny Coating} \\ \hline
{\tiny SACK010-0213  ({\bf A})}& {\tiny16.51} &{\tiny 77.7} & {\tiny Al+SiO$_2$+HfO$_2$+SiO$_2$ } \\ \hline
{\tiny SACPNG01-0213 ({\bf B})}&  {\tiny 16.37} & {\tiny 65.3}& {\tiny Al+SiO$_2$+HfO$_2$+SiO$_2$}\\ \hline
\end{tabular}
\caption{Irfu-CEA/Kerdry  MST mirror facets characteristics.}
\label{table:mirrors}
\end{center}
\end{table}

The two MST mirrors are not identical but they weight approximately 25 kg, being  both completely watertight. Facets for the MST, built using other technologies, are expected to be provided by other groups to perform the same test at SAC.  

\subsection{Optical and mechanical requirements to test}\label{sec:req}

The important optical properties that should be stable in real observation conditions are: the mirror focal length, the ability to focus determined by the size of the Point Spread Function (PSF) and the reflectivity (global and local). The focal length and the PSF can be evaluated using a simplified ``2f'' system (see section \ref{sec:facility}). In the case of the reflectivity, only the local reflectivity measurement could be easily implemented at the site, using an adapted commercial spectrometer. 
The optical specifications should be fulfilled within the operational temperature range measured at the SAC site and the mirrors shall not suffer any damage or irreversible change of optical properties from temperature variations within -25$^\circ$C to +60$^\circ$C. 
The global reflectivity, defined as the percentage of reflected light that is focused by the mirror inside a circular area with an integration diameter of 1 mrad, shall not change more than 3$\%$. If the mirror shape is affected by the atmospheric variations, this will impact on the PSF and finally on the global reflectivity.  
From the mechanical point of view, the mirrors should resist winds loads of $\sim$120 km/h,  dust abrasion and eventual impacts, without suffering permanent damage or change. This would be detected as variations in the focal length or the reflectivity of the mirrors. 

\section{Site description} \label{sec:site}
\begin{figure}[h]
  \centering
  \includegraphics[width=0.5\textwidth] {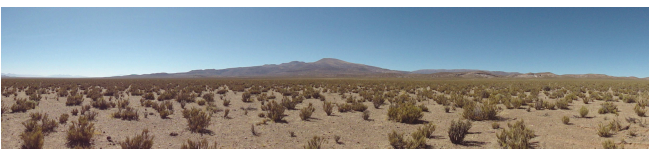} 
  \caption{Panoramic view of the San Antonio de los Cobres site.}
  \label{fig:SAC}
 \end{figure}

The San Antonio de los Cobres site is located at Lat. 24$^\circ$ 02' 42.7'' S and Lon. 66$^\circ$ 14' 05.8'' W, in the Province of Salta, Argentina. Altitude of the site is 3600 meters a.~s.~l. It is a flat area of approximately 4 km $\times$ 5 km surrounded by hills of hundreds of meters altitude above the plateau (see Fig. \ref{fig:SAC}). In the region a semi-arid continental climate dominates, with high temperature fluctuations between day and night. Since 2010, several instruments have been installed to monitor the general atmospheric conditions (temperature, humidity, wind speed, etc.), the night sky background and the cloud coverage (see \cite{bib:SiteInst-1} for details). 
 
\section{Test facility description}\label{sec:facility}

Inside the fenced area at the SAC site, it was possible to place only four MST mirrors in a suitable position for the tests. The system consists of four dedicated structures to hold the mirrors in three different positions: {\it parking} (mirrors facing down), {\it observation} (mirrors facing up, at -45$^\circ$ from the zenith) and {\it test} (mirrors in vertical position facing the horizon). During daytime the mirrors remain at {\it parking} position. After sunset, they move to {\it observation}, coming back to {\it parking} at sunrise. The {\it test} position is used when the optical properties are tested using a portable ``2f'' system. 

\begin{figure}[h]
  \centering
  \includegraphics[width=0.5\textwidth] {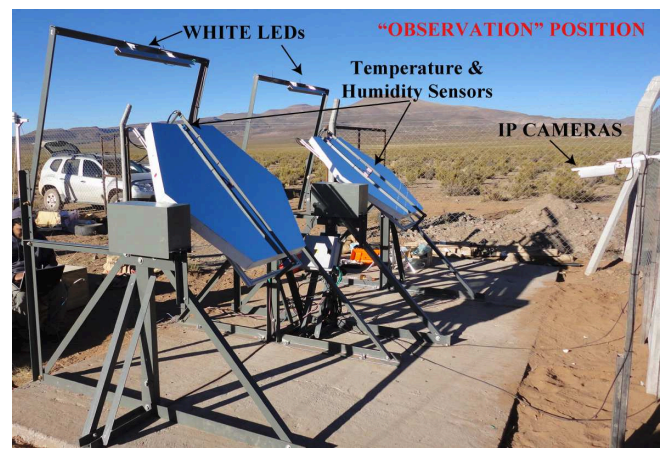}
  \caption{Mirrors test facility at San Antonio de los Cobres, Province of Salta, Argentina.}
  \label{fig:facility}
 \end{figure}

Each mirror is viewed by an IP camera (Ubiquiti AirCam \footnote{http:$\/\/$www.ubnt.com}) and images of the mirrors are taken every 10 minutes. In order to distinguish the fog or ice formation on the mirror surfaces when it happens, they are illuminated by high intensity white LED (SMD cold white 5060 \footnote{http://dled.com.ar/leds-5060-exterior}) during the taking of photos (see Fig.~\ref{fig:facility}). The cameras and the movement of the mirrors are controlled by a SBC TS-7260 \footnote{http://www.embeddedarm.com/documentation/ts-7260-datasheet.pdf}. The images and sensors data (temperature and humidity) are stored in a 32G pen-drive. A power control system has been developed for controlling the high intensity LED with the SBC.

Two pairs of sensors for temperature and relative humidity (RTD\footnote{{\it Resistance Temperature Detectors} (RTD, Honeywell HEL777AUO)} and HIH-400\footnote{http://www.phanderson.com/hih-4000.pdf}, respectively) are placed on the surface of both mirrors, and one additional pair measures the air temperature and humidity. The data is sent via the RS-232 port when the SBC requests it.  

\begin{figure}[h]
  \centering
  \includegraphics[width=0.25\textwidth] {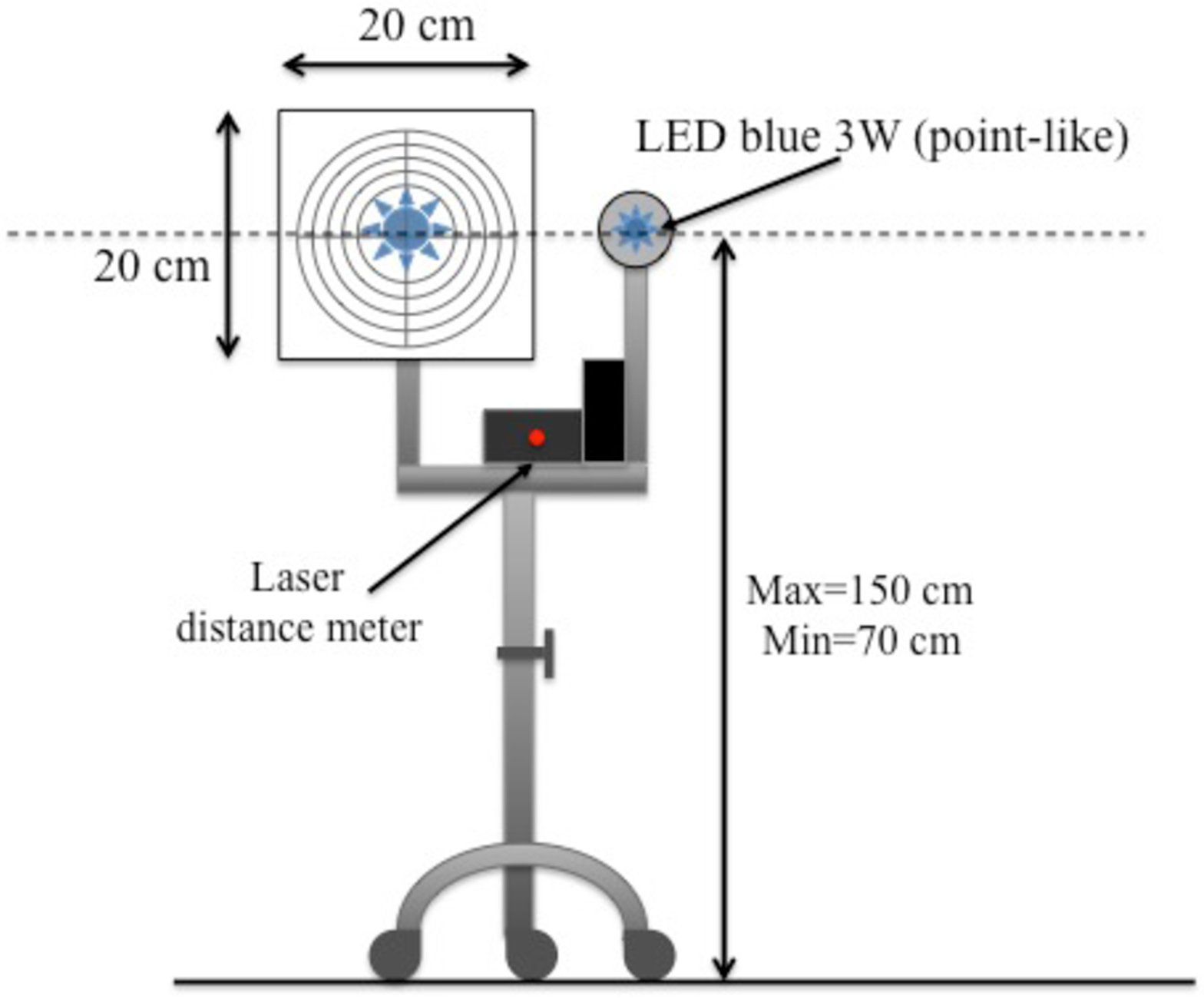}
  \includegraphics[width=0.19\textwidth] {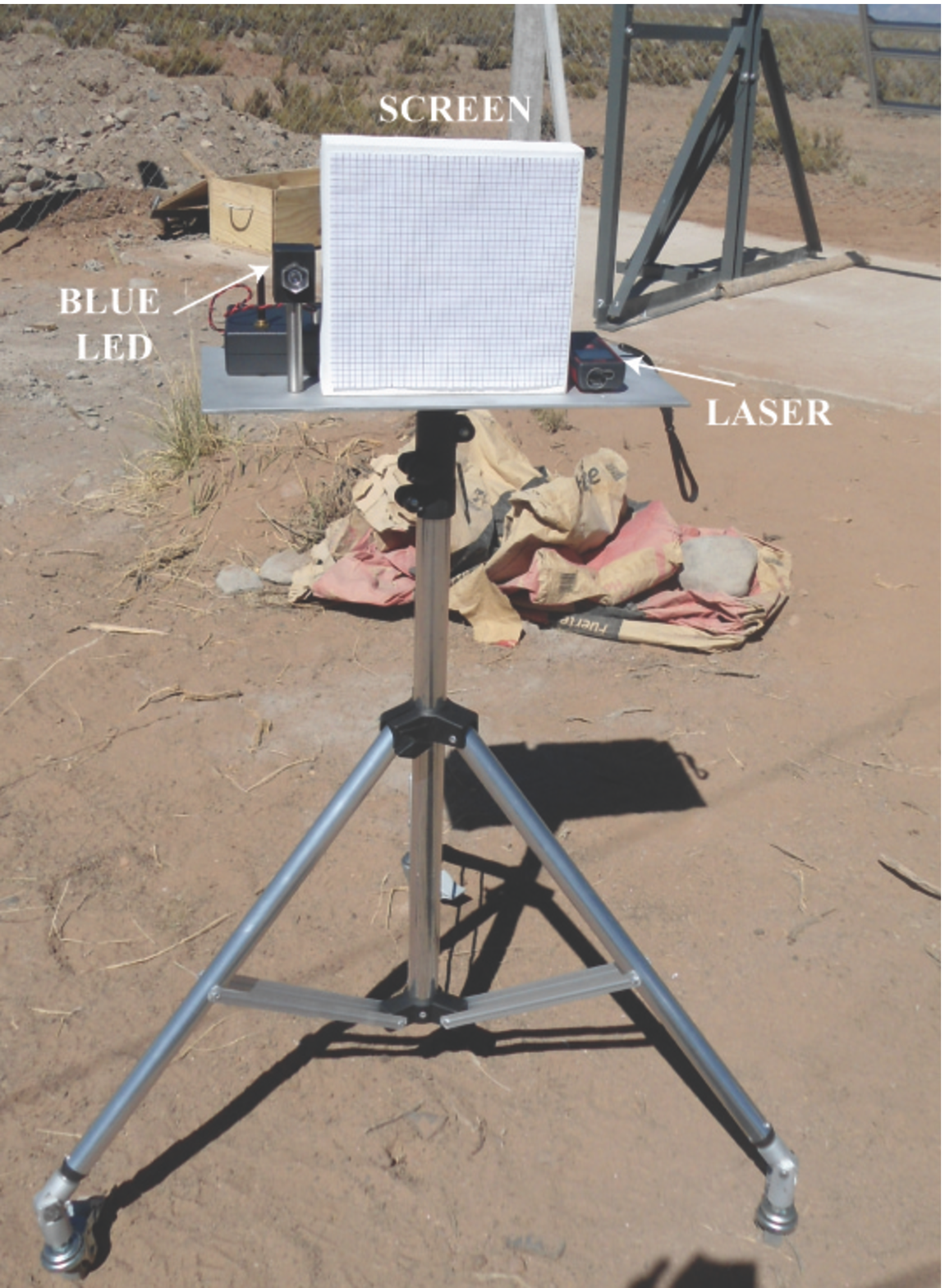}
  \caption{{\it Left}: Portable {\it 2f} system schematics. {\it Right}: System used at SAC for the outdoor measurements.}
  \label{fig:sistema2f}
 \end{figure}

In order to evaluate the evolution of the optical properties we have developed a portable {\it 2f} ({\it f}~=~mirror focal length) system following the specifications described in \cite{bib:Saclay}. With this system the mirror is uniformly illuminated by a light source placed at twice the mirror focal length (2f) and close to its optical axis. The light source should be point-like (much smaller than the mirror PSF). The light reflected by the mirror will ideally produce at {\it 2f} an inverted 1:1 scale image of the source. The spread of this image is twice the mirror PSF. 
In our case, we use a 3 Watts blue LED to illuminate the mirror and the image formed at the screen placed next to the light source is captured with a commercial camera (see right panel of Fig. \ref{fig:sistema2f}). The screen and the light source are attached to a tripod which allows to adjust the distance of the system to the mirrors (see left panel of Fig. \ref{fig:sistema2f}). 
 
In principle, the focal length of the mirrors can be determined by scanning the spot size with the distance to the mirror. However, the lack of uniformity on the ground (presence of a slope and high density of bushes) made very difficult a smooth movement of the system, preventing a successful scan of the image size. Some improvements on the portable {\it 2f} system will be implemented in the next months in order to avoid these difficulties. These will consist basically on a more stable support, with a better light collimator and a fixation for the camera for keeping the same relative position between the screen and the camera in each measurement. 

\section{Preliminary results}\label{sec:results}

\subsection{Condensation study}
\begin{figure}[!h]
  \centering
  \includegraphics[width=0.24\textwidth] {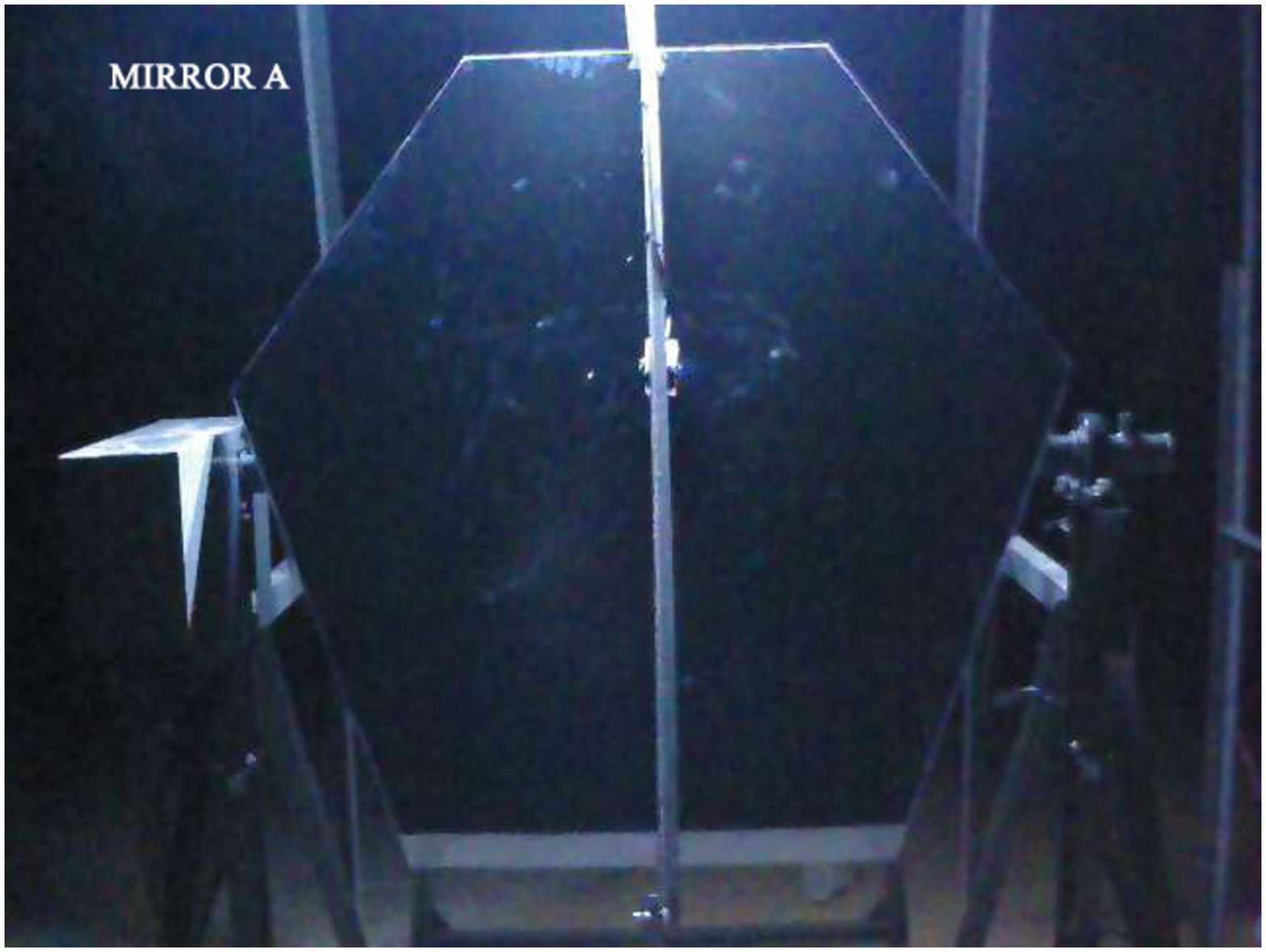}
  \includegraphics[width=0.24\textwidth] {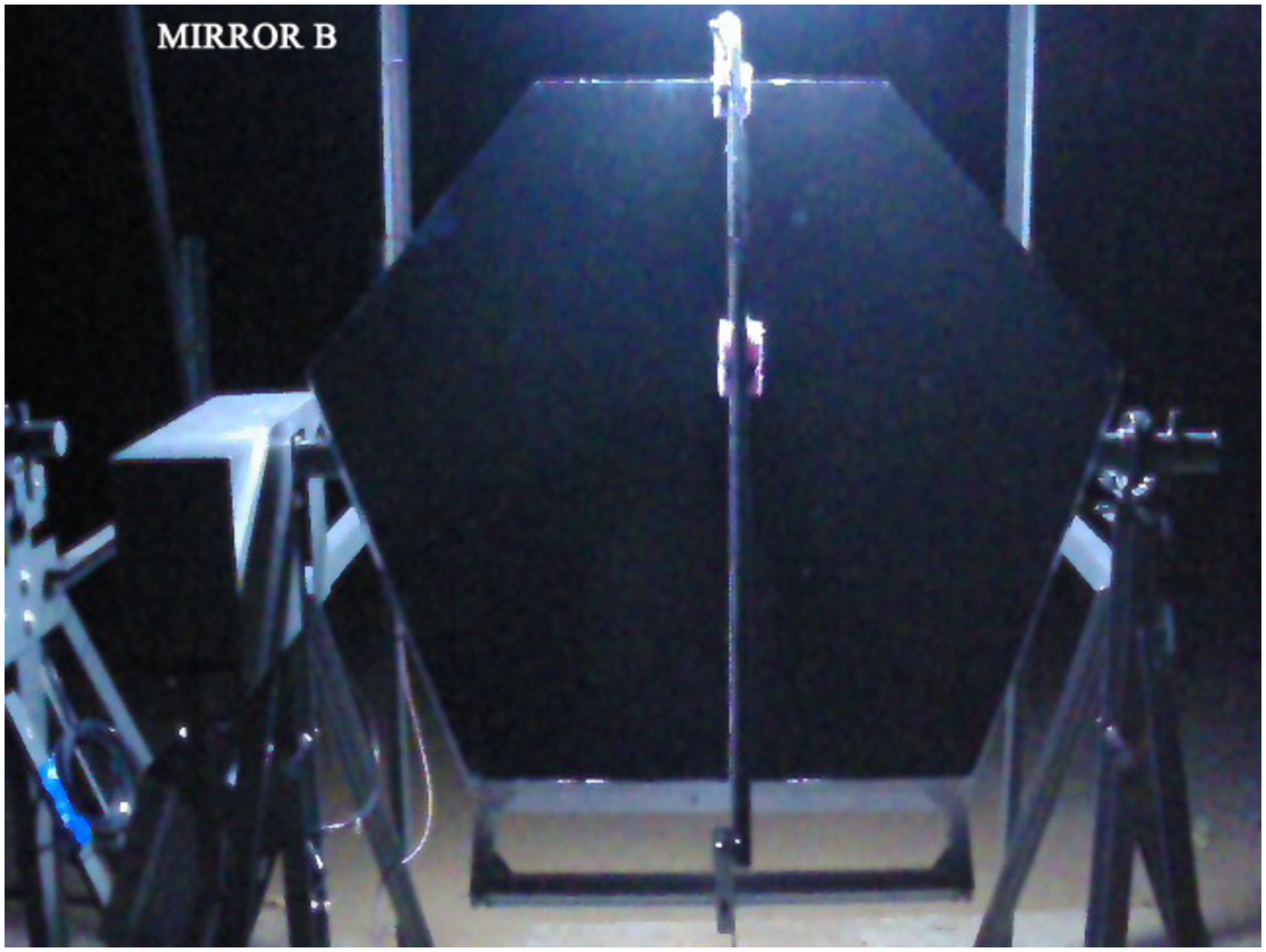}
  \caption{Mirrors surface captured by the IP Camera during the night of May 12th, 2013. No features due to icing or condensation are seen in the images. {\it Left}: Mirror SACK010-0213 (A). {\it Right}: Mirror SACPNG01-0213 (B).}
  \label{fig:foto}
 \end{figure}

\begin{figure}[h]
  \centering
  \includegraphics[width=0.48\textwidth] {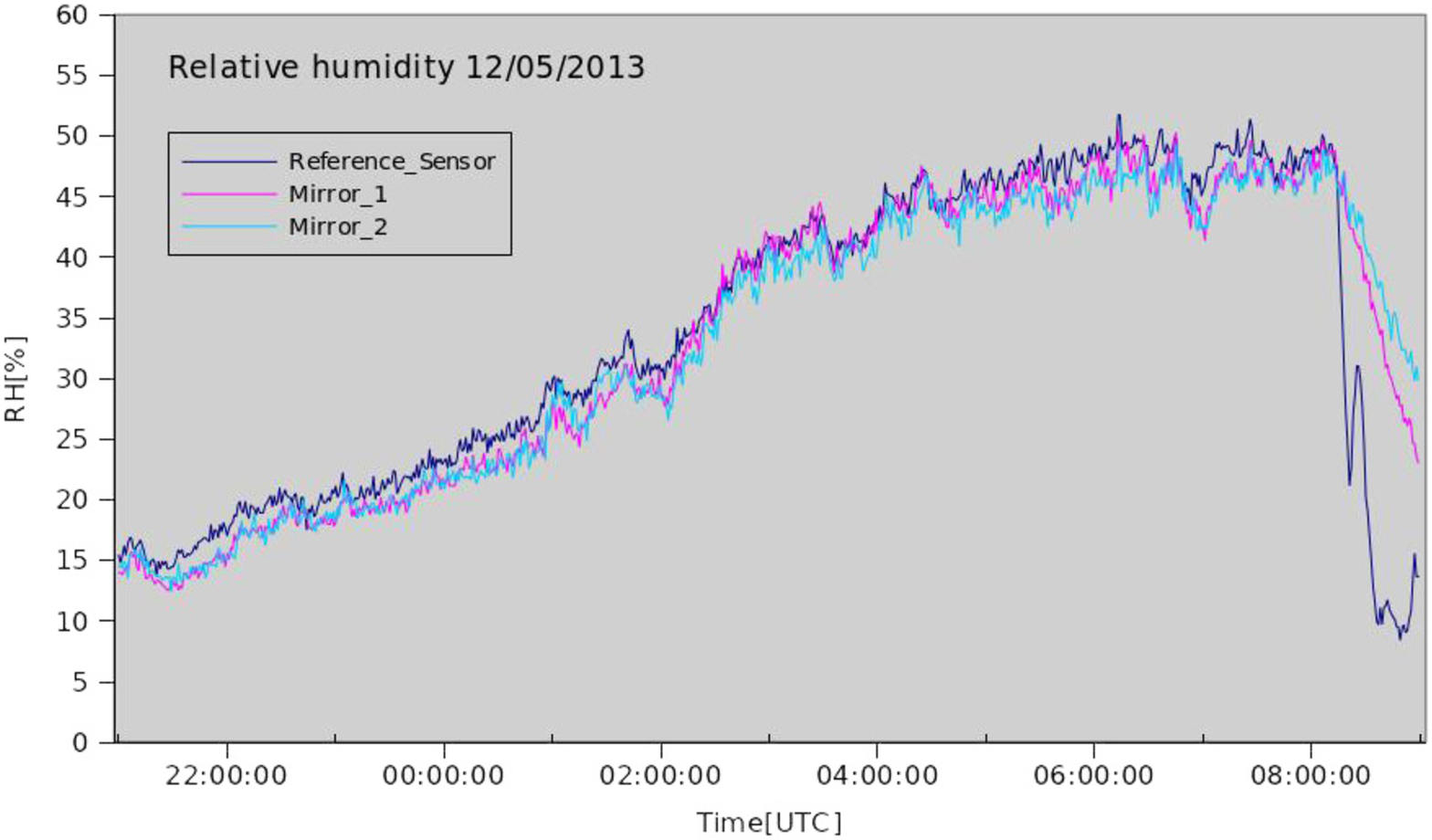} 
  \includegraphics[width=0.48\textwidth] {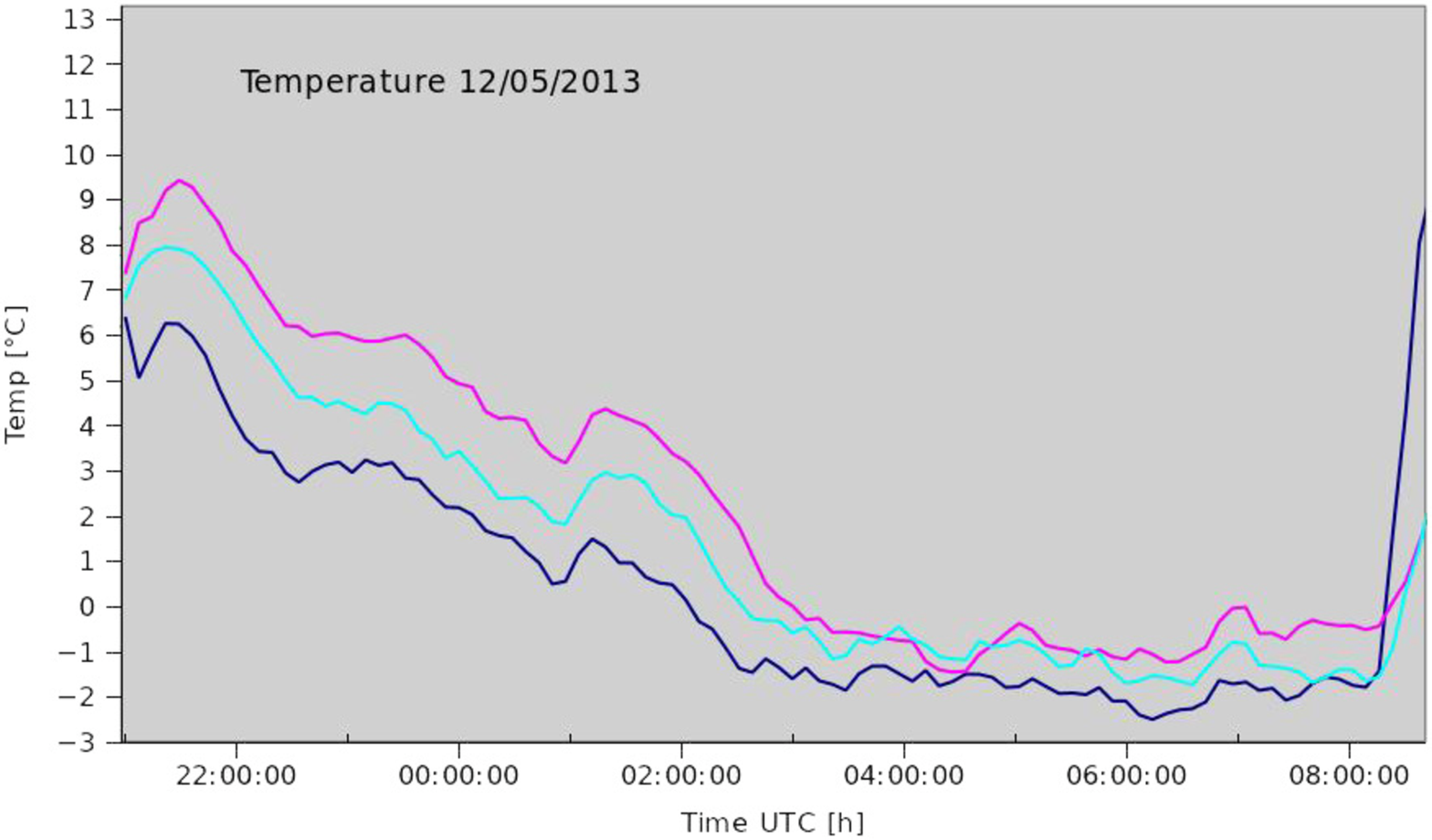} 
  \caption{Humidity and temperature evolution during May 12, 2013 night. {\it Top}: Relative humidity registered by the sensors placed near the mirrors reflective surface (magenta and cyan lines) compared to the one taken by the {\it reference} sensor (blue line). {\it Bottom}: Temperature registered by the sensors placed together with the humidity ones. The same color scheme is used in this plot. }
  \label{fig:HyT}
 \end{figure}

The system is working since May 10th, 2013. The first images of the mirrors (see Fig.~\ref{fig:foto}) show that no condensation is produced on the mirror surfaces during the first night of operation. The values of humidity and temperature registered by the sensors for the same night, can be seen in Fig.\ref{fig:HyT}. 

They are compared to a {\it reference}  pair of sensors which was placed nearby the mirrors. We can see that the relative humidity maximum value is less than 50$\%$ with a temperature minimum greater than -3$^\circ$C. 

\subsection{Optical parameters}

\begin{figure}[h]
  \centering
  \includegraphics[width=0.22\textwidth] {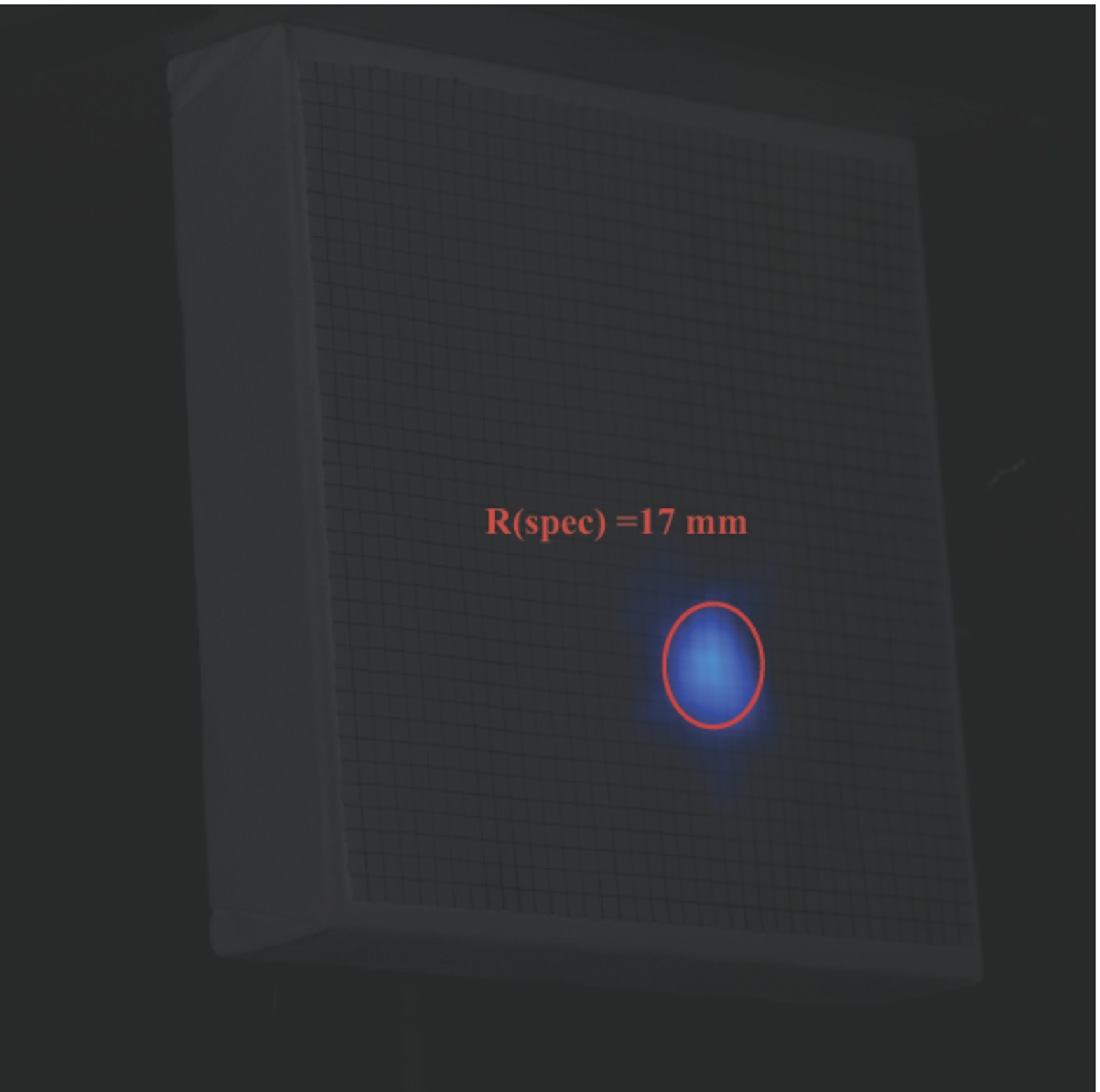}
 \includegraphics[width=0.25\textwidth] {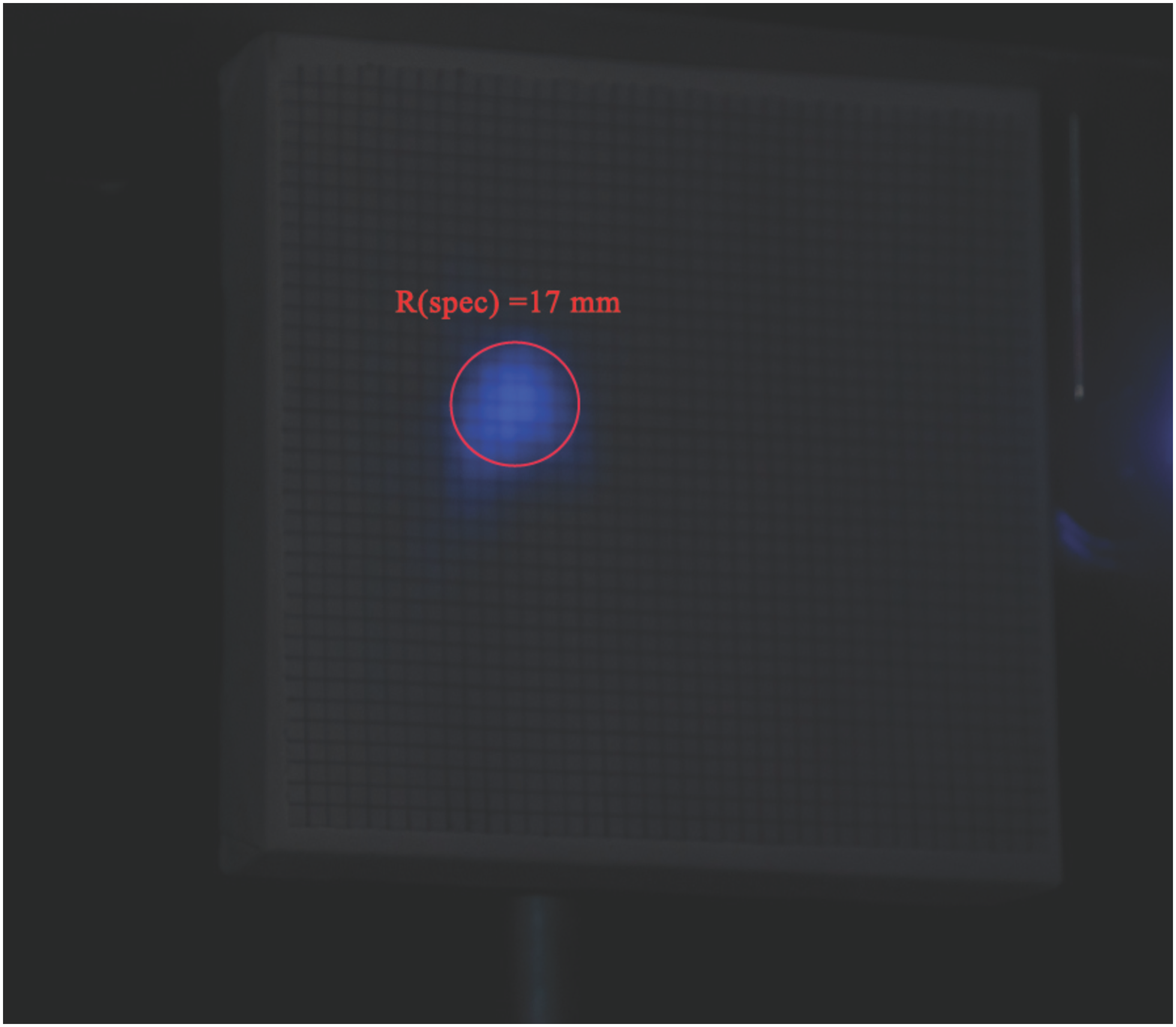}
  \caption{{\it Left}:  Image produced at twice the focal length of the mirror A.{\it Right}: Image produced at twice the focal length of the mirror B.}
  \label{fig:PSF}
 \end{figure}

Even though the {\it 2f} system shall be improved soon, we were able to capture the spot produced by the two mirrors and estimate their focal length, finding the smallest spot size. This is shown in Fig.~\ref{fig:PSF}. These are very preliminary results but the minimal image size was found at $\sim$32.90~m~$\pm$~0.10~m for mirror A and $\sim$32.5~m~$\pm$~0.10~m  for mirror B.  

\section{Conclusions}\label{sec:conc}

The first outdoor test facility for mirrors has been installed at the northern Argentinean candidate site for CTA (SAC, Province of Salta). This facility consist of a dedicated structure to securely fix the mirrors together with a remote positioning system which allows to move the mirrors from a {\it parking} position during daytime to an {\it observation} position at night. The first two MST mirrors provided by the Irfu-CEA group and the Kerdry company have been already installed at the site. Temperature and humidity sensors have also been installed near the mirror surfaces. The behavior of the mirrors regarding fogging and icing is monitored by means of pictures taken by an IP camera every ten minutes during night time. Images of the mirrors are being taken and the weather conditions are being monitored. The very first results related to fogging phenomena are encouraging but more statistics and a longer time coverage is needed. The optical stability will also be tested for the next 6 months using an improved version of the portable 2f system. Other mirrors built with different technologies are also intended to be installed at the SAC site in the following months.

\vspace*{0.5cm}
\footnotesize{{\bf Acknowledgment:} {We gratefully acknowledge support from the following agencies and organizations:
Ministerio de Ciencia, Tecnolog\'ia e Innovaci\'on Productiva (MinCyT),
Comisi\'on Nacional de Energ\'ia At\'omica (CNEA) and Consejo Nacional  de
Investigaciones Cient\'ificas y T\'ecnicas (CONICET) Argentina; State Committee
of Science of Armenia; Ministry for Research, CNRS-INSU and CNRS-IN2P3,
Irfu-CEA, ANR, France; Max Planck Society, BMBF, DESY, Helmholtz Association,
Germany; MIUR, Italy; Netherlands Research School for Astronomy (NOVA),
Netherlands Organization for Scientific Research (NWO); Ministry of Science and
Higher Education and the National Centre for Research and Development, Poland;
MICINN support through the National R+D+I, CDTI funding plans and the CPAN and
MultiDark Consolider-Ingenio 2010 programme, Spain; Swedish Research Council,
Royal Swedish Academy of Sciences financed, Sweden; Swiss National Science
Foundation (SNSF), Switzerland; Leverhulme Trust, Royal Society, Science and
Technologies Facilities Council, Durham University, UK; National Science
Foundation, Department of Energy, Argonne National Laboratory, University of
California, University of Chicago, Iowa State University, Institute for Nuclear
and Particle Astrophysics (INPAC-MRPI program), Washington University McDonnell 
Center for the Space Sciences, USA. The research leading to these results has 
received funding from the European Union's Seventh Framework Programme 
([FP7/2007-2013] [FP7/2007-2011]) under grant agreement nå¡ 262053.}}

\end{document}